\documentclass[pra,aps,showpacs]{revtex4}
\usepackage{amsmath,amssymb}
\usepackage{graphicx}

\newcommand{\la}{\langle}
\newcommand{\ra}{\rangle}

\begin{document}

\title{Dynamics of heuristic optimization algorithms on random graphs}
\author{Martin Weigt}
\affiliation{Institute for Theoretical Physics,
University of G\"ottingen, Bunsenstr. 9, 37073 G\"ottingen, Germany,\\
E-mail: {\tt weigt@theorie.physik.uni-goettingen.de}
}

\date{\today}

\begin{abstract}
In this paper, the dynamics of heuristic algorithms for constructing
small vertex covers (or independent sets) of finite-connectivity
random graphs is analysed. In every algorithmic step, a vertex is
chosen with respect to its vertex degree. This vertex, and some
environment of it, is covered and removed from the graph. This graph
reduction process can be described as a Markovian dynamics in the
space of random graphs of arbitrary degree distribution. We discuss
some solvable cases, including algorithms already analysed using
different techniques, and develop approximation schemes for more
complicated cases. The approximations are corroborated by numerical
simulations.
\end{abstract}


\maketitle


\section{Introduction}\label{sec:intro}

Many questions of practical or scientific interest are based on
combinatorial optimization problems whose numerical solution requires 
time resources growing exponentially with the system size, or more
precisely, with the number of binary variables needed to encode the
problem. These include examples like planning and scheduling problems
in various real-world applications, optimization of chip design,
cryptografic systems in computer science, or glassy systems and random 
structures in physics. All these problems are characterized by a
non-trivial cost function, or energy, which has to be minimized over 
a large set of discrete degrees of freedom.

The hardest of these optimization problems are collected in a class
called {\em NP-hard} \cite{GaJo}. Hardness refers in this context to
the exponential growth of the computational resolution time which is
observed for all known numerical algorithms. Despite an extremely
large number of known NP-hard problems, and numerous approaches to
solve them, no algorithms could be constructed by now which are able
to solve such a problem in a time growing only polynomially with the
system size. This point supports the widespread conjecture, that no 
such effective algorithms are constructible.

The numerical search for globally optimal solutions is thus restricted
to relatively small systems. Once one has to solve larger systems,
good polynomial-time algorithms are needed which construct low-cost
configurations. These are not guaranteed to be optimal, but in many
cases they can serve as reasonable approximations \cite{note:ptas}. 
Many of these algorithms are based on heuristic considerations,
e.g. on expected correlations between local structures of the specific
problem instance and its optimal solutions. Exploiting these
correlations can largely improve the performance of a heuristic 
procedure.

Here we are using the example of vertex covers (VC) on random graphs.
It belongs to the basic NP-hard problems \cite{GaJo} and can be
considered as a prototype optimization problem over a random
structure.  In every step of the presented heuristic algorithms, a
vertex and possibly some environment is chosen randomly and covered
locally optimal. The local structure of the graph, in particular its
vertex-degree distribution (distribution of co-ordination numbers),
can be exploited: Vertices of high degree are more likely to be
covered, those of small degree are more likely to remain uncovered.
We are using random graphs in order to get some information about the
{\it typical-case} behaviour of the algorithm. This is to be
contrasted with the worst-case picture used in the traditional theory
of computational complexity \cite{GaJo}.

The study of heuristic algorithms is also interesting from a more
theoretical point of view. Many randomized optimization or decision
problems show characteristic phase transition, when the parameters of
the randomness are tuned, see e.g. the special issues \cite{AI,TCS}
for an overview. The analysis of algorithms is frequently used in
theoretical computer science to construct bounds for these phase
transitions, for some examples see e.g. \cite{Ga,Wo,PiSpWo,Fra,Ac} and
references therein. A different approach to these transitions is given
by applying techniques from equilibrium statistical mechanics, as was
done successfully for some of the fundamental combinatorial problems
like 3-satisfiability \cite{MoZe,nature}, number partitioning
\cite{Me}, or also vertex cover \cite{WeHa1,WeHa2}.

This paper is organized as follows: In section \ref{sec:model}, we
introduce the definition of vertex covers and independent sets,
present the algorithms and review some important facts on random
graphs. The general dynamical equations for the graph evolution
process are developed in section \ref{sec:dynamics}, and they are
solved for some cases in section \ref{sec:linear}. Section
\ref{sec:nonlinear} is dedicated to approximations of the dynamical
behaviour of cases which could not be solved exactly. In section
\ref{sec:leaf}, we present how the leaf-removal algorithm of Bauer
and Golinelli \cite{BaGo1,BaGo2} fits into the presented scheme. It is
also generalized to cases where the original algorithm fails to
construct a vertex cover. The last section finally summarizes the
results and gives an outlook to possible extensions.


\section{Model and algorithms}\label{sec:model}

\subsection{Vertex cover and related problems}
\label{sec:vc}

Let us start with the definition of vertex covers \cite{GaJo}:

Take any graph $G=(V,E)$ with $N$ vertices $i\in\{1,...,N\}$
and $M$ undirected edges $\{i,j\}\in E\subset V\times V$. 
A {\it vertex cover} (VC) is a subset $U \subset V$ of vertices 
such that for every edge $\{i,j\}\in E$ there is at least one of 
its endpoints $i$ or $j$ in $U$:
\begin{equation}
  \label{eq:defvc}
  U \subset V \mbox{ is VC}\qquad  \leftrightarrow \qquad
  \forall \{i,j\}\in E : \quad i\in U\ \vee\ j \in U
\end{equation}
We call the vertices in $U$ covered ({\it cov}), whereas the
vertices in its complement $V\setminus U$ are called uncovered ({\it
  uncov}). The definition of a vertex cover implies therefore that
every edge has at least one covered end-point.

The full vertex set $V$ is of course a trivial vertex cover of any
graph $G$. In this case, all edges have two covered end-points, and at
least some of the vertices can be set {\it uncov} without uncovering
any edge. The corresponding {\it optimization problem} consists
in finding a vertex cover of smallest cardinality. This problem
belongs, according to the standard book by Garey and Johnson
\cite{GaJo}, to the basic NP-hard optimization problems. Therefore, it 
is expected to require a solution time growing exponentially in $N$
and $M$. The numerical solvability is consequently restricted to
relatively small graphs.

VC is related to other well-known and widely used NP-hard
problems. The first one is the {\it independent set} (IS) problem.
An IS is a subset $S\subset V$ of vertices such that  for
all $i,j\in S$ we have $\{i,j\}\notin E$. So $V\setminus S$ is obviously 
a VC for every IS $S$, and every maximal IS is the complement of a 
minimal VC. The {\it independence number}, defined as the maximal  
cardinality max$(|S|)$ of all ISs, is consequently given by 
$N-\min_{\mbox{VC } U} |U|$.

A {\it clique} is a fully connected subgraph. So, if the vertex subset
$S\subset V$ is an IS in $G=(V,E)$, it is a clique in the
complementary graph $\overline{G}=(V,V\times V \setminus E)$. Finding
the largest clique in one graph is equivalent to finding the largest
independent set in the complementary graph.

There is also a physical problem which is equivalent to vertex cover, 
or more obviously to the independent set problem. Imagine the graph
$G$ to be a lattice, and pack hard spheres of chemical radius 1 onto
the vertices. Then, once a vertex is occupied by such a particle, all 
neighbouring vertices have to be empty. This is exactly what defines
an IS. The vertices which are not occupied by spheres thus form a VC.
This equivalence provides the basis of the statistical mechanics'
approach to minimal vertex covers on random graphs \cite{WeHa3}.

\subsection{Heuristic algorithms for constructing small vertex covers}
\label{sec:algo}

As already mentioned, the construction of a minimal vertex cover is
NP-hard, thus requiring exponential time resources. It is therefore
reasonable to develop good approximation algorithms running in
polynomial time. Here we describe a class of linear time heuristic
algorithms which are able to produce small, but in general suboptimal
VCs.

In this algorithm, vertices are sequentially assigned the values {\it
cov} and {\it uncov} until the full graph is covered. An assigned
value is not changed any more. This can be interpreted as a graph
reduction process: Once a vertex is covered, it can be removed from
the graph, together with all incident edges. If a vertex is set {\it
uncov}, all its neighbours have to be covered in order to cover the
graph. The central vertex, its neighbours and all covered edges can
again be removed from the graph. The graph thus becomes smaller and
smaller, until no edges are left. The size of the resulting vertex
cover crucially depends on the order of the vertex selection and the
decision to cover/uncover the selected vertex. The main heuristic idea
is simple: A vertex of high degree is more likely to be covered, a
vertex of small degree is more likely to be uncovered
\cite{WeHa2,WeHa3}. The simplest local information, namely the vertex
degree, is thus correlated to the structure of small vertex covers,
and can be exploited algorithmically.

This is done in the following way: Given an initially uncovered graph
$G=(V,E)$ and a non-negative integer $k$, which we call the {\it
depth} of the algorithm. In every algorithmic step a vertex $i$ is
chosen randomly, and its nearest, 2nd-nearest... and $k$th-nearest
neighbours are selected. All these vertices, together with all edges
connecting two of it, form the induced subgraph
$G^{(k)}(i)=(V^{(k)}(i),E^{(k)}(i))$. This subgraph can be efficiently 
covered such that all $k$th nearest neighbours are set to {\it cov}. 
For details see the algorithm presented below. The full subgraphs,
together with all edges connecting it to other vertices, is deleted 
from $G$.

For locally tree-like graphs, as e.g. random graphs or Bethe lattices,
the vertex covers of $G^{(k)}(i)$ are especially simple: The $k$th
neighbours of $i$ are covered, the $(k-1)$st are uncovered, the
$(k-2)$nd are covered again, and so on, until $i$ itself is
covered (uncovered) for even (odd) depth $k$ of the algorithm.

The main heuristic concerns now the question, how the central vertex
$i$ is selected. It exploits the above-mentioned correlation between
vertex degree and covering state of an arbitrary vertex. For even
depth $k$, the central vertex is covered by the above procedure 
\cite{note1}. So it is useful to choose more
frequently vertices of high degree in the subgraph remaining after
elimination of already considered vertices. We therefore select a
vertex of degree $d$ with some weight $w_d$ which is a monotonously
growing function of $d$.

In the case of odd $k$, the central vertex is uncovered \cite{note1}.
Here, the selection weight $w_d$ for vertices has to be a
monotonously decreasing function of the vertex degree $d$.

The algorithm is summarized as follows, inputs are the graph
$G=(V,E)$, the non-negative integer depth $k$, a mapping $m : \{1,...,N\}
\to \{free,cov,uncov\}$, which is constantly set to $m(i)\equiv free$
initially, and the positive weight function $w:\mathbb N \to \mathbb
R_+$:

\newlength{\tablen}
\settowidth{\tablen}{xxx}
\newcommand{\tabspace}{\hspace*{\tablen}}
\begin{tabbing}
\tabspace \= \tabspace \= \tabspace \= \tabspace \= \tabspace \=
\tabspace \= \kill
{\bf procedure} heuristic-VC($G, k, m, w$)\\
{\bf begin}\\
\> {\bf if} $E = \emptyset$ {\bf then}\\
\>\> {\bf stop};\hskip 1cm 
     \{all edges are covered by vertices with $m(i)=cov$\}\\
\> Select a vertex $i\in V$ of current degree $d(i)$ randomly with 
   weight $w_{d(i)}$;\\
\> {\bf if} $d(i) = 0$\\
\>\> {\bf begin}\\
\>\>\> $m(i) := uncov$;\\
\>\>\> $V := V\setminus \{i\}$;\\
\>\>\> heuristic-VC($G, k, m, w$);\\
\>\> {\bf end};\\
\> {\bf else}\\
\>\> {\bf begin}\\
\>\>\> $V^{(0)}(i) := \{i\}$;\\
\>\>\> {\bf for $\kappa$ from $1$ to $k$}\\
\>\>\>\> $V^{(\kappa)}(i) := V^{(\kappa-1)}(i) \cup 
     \{ \kappa\mbox{th nearest neighbours of } i\}$;\\
\>\>\> $E^{(k+1)}(i) := \{\ \{i,j\} \in E\ |\ i \in V^{(k)}(i)\ \vee\  
   j \in V^{(k)}(i)\ \}$;\\
\>\>\> $V := V \setminus V^{(k)}(i)$;\\
\>\>\> $E := E \setminus E^{(k+1)}(i)$;\\
\>\>\> $m(j) := cov$ for all $k$th neighbours of $i$;\\
\>\>\> {\bf while} $V^{(\kappa-1)}(i) \neq \emptyset$\\
\>\>\>\> {\bf begin}\\
\>\>\>\>\> Select a vertex $j\in V^{(\kappa-1)}(i)$ of maximal distance
         from $i$;\\ 
\>\>\>\>\> $m(j) := uncov$;\\
\>\>\>\>\> $m(l) := cov$ for all neighbours $l\in V^{(\kappa-1)}(i)$ of
         $j$;\\ 
\>\>\>\>\> $V^{(\kappa-1)}(i) := V^{(\kappa-1)}(i) \setminus \{j, 
         \mbox{neighbours of } j\}$;\\
\>\>\>\> {\bf end};\hskip 1cm  \{subgraph covered and removed from $G$\}\\
\>\>\> heuristic-VC($G, k, m, w$);\\
\>\> {\bf end};\\
{\bf end};\\
\end{tabbing}

Please note that the degree of free vertices may change whenever $G$
is reduced. The algorithm always considers the current degree in the
reduced subgraph, which equals the number of uncovered incident edges.
The algorithm therefore defines a Markov process.

Some special cases of the algorithm where already considered for
finite connectivity random graphs: In \cite{WeHa4}, the case $k=0$ and
$w_d=1$ was included into a complete backtracking algorithm. The upper
bound for the minimal vertex covers constructed in this way was rather
poor, but it will be improved in sections \ref{sec:linear} and
\ref{sec:nonlinear} by using better $w_d$. The case of $k=1, \ w_d=1$
was analyzed in \cite{Ga} using a different technique. Whereas being
also quite unsatisfactory for small and intermediate average graph
connectivities $c_0=2M/N$, this algorithm correctly reproduces the
leading asymptotic behaviour for large $c_0$. In \cite{BaGo2}, very
surprising results where obtained for $k=1, \ w_d=\delta_{d,1}$, where
only vertices of degree 1 (leafs) are selected and uncovered, their
neighbour is covered, and both are removed from the graph. This
algorithm is able to cover almost all edges for $c<e$, thus producing
a minimal vertex cover, but it stops for higher connectivities if no
vertices of degree 1 are left, even if an extensive number of edges
remains uncovered. These two results suggest a promising
generalization for the case $k=1$: If we choose $w_1 \gg w_d >0$ for
all $d>1$, this algorithm will work nearly as well as the leaf
removal procedure for small connectivities, but it will also give the
correct asymptotic behaviour for large $c_0$. The extreme case of
choosing always a vertex of smallest current degree will work best on
random graphs, but it goes beyond the analysis presented in this
paper.

\subsection{Random graphs of arbitrary degree distribution}
\label{sec:randomgraph}

In order to gain some insight into the typical behaviour of this
algorithm, we apply it to random graphs. This subsection is dedicated
to summarizing some interesting known results about these graphs, as
far as they are important for our analysis. As can be expected from 
the algorithm presented above, we will concentrate our attention to 
the distribution of vertex degrees. For a complete presentation see
\cite{Bo}.

The original idea \cite{ErRe} is to assign an equal probability to all
graphs having the same numbers $N$ of vertices and $M$ of edges. A
random graph $G_{N,p}$, with $0\leq p\leq 1$, is constructed in the
following way: The vertex set is chosen to be $V=\{1,2,...,N\}$. For
all vertex pairs $i,j\in V,\ i<j,$ an edges is included into $E$ with
probability $p$. The two vertices remain disconnected by a direct edge
with probability $1-p$. This graph has on average $M=p {N \choose 2}$
edges, its mean vertex degree equals $c_0=(N-1)p$.

The most interesting case for vertex covers are graphs of finite average 
connectivity, i.e. the average vertex degree $c_0=2M/N$ stays finite 
in the thermodynamic limit $N\to\infty$. In the above language, we
have to fix $p=\frac{c_0}{N-1}$. The resulting degree distribution  
is far from uniform. For $N\gg 1$, a randomly chosen vertex has degree 
$d$  with probability
\begin{equation}
  \label{eq:poisson}
  p_d = e^{-c_0} \frac{c_0^d}{d!}\ ,
\end{equation}
i.e. the degree distribution approaches a Poissonian in the
thermodynamic limit. In our analysis we also need the probability of 
finding a vertex of degree $d$ by following an arbitrary edge. It is 
obviously proportional to $d\ p_d$, due to normalization we have
\begin{equation}
  \label{eq:neighbourpoisson}
  p^{(1)}_d = e^{-c_0} \frac{c_0^{d-1}}{(d-1)!}\ .
\end{equation}
We thus find a Poissonian distribution of $d-1$. The average degree of
vertices reached by following an edge equals $c_0+1$, so there are on
average $c_0$ additional edges.

Random graphs undergo a percolation transition at average vertex
degree $c_0=1$. Below this threshold, the graph consists of an
extensive number of small connected components, each containing up to
$O(\ln N)$ vertices.  For $c_0>1$, the number of small connected
components of $G_{N,c/(N-1)}$ is still extensive, but there is also
one macroscopic connected component of $O(N)$ vertices. This giant
component grows with increasing $c_0$, and exponentially approaches
size $N$ when $c_0$ becomes large.

The concept of random graphs was recently generalized to random graphs
of arbitrary degree distribution \cite{MoRe,Ne}. There, every graph of
a given distribution $p_d$ (not necessarily a Poissonian) is assigned
the same probability. These graphs can be easily generated: For all
vertices $i=1,...,N$, a degree $d(i)$ is drawn randomly from $p_d$. If
$\sum_i d(i)$ is even, we continue, if not, we repeat the above
procedure. Then a large vertex set is created, containing every vertex
$i$ exactly $d(i)$ times. Now we sequentially select pairs of vertices
and add these to the edge set $E$, excluding only
self-connections. Following again an edge, the reached vertex has
degree $d$ with probability $p^{(1)}_d = d\ p_d/c_0$, with $c_0=\sum_k
k\ p_k$ denoting the average degree.

\subsection{Minimal vertex covers on random graphs}
\label{sec:minvc}

Before analysing the behaviour of the presented heuristics, we will
give a short overview over known properties of minimal vertex covers
on random graphs, see \cite{WeHa3} for the original presentation.

The analysis there was carried out for random graphs $G_{N,c_0/(N-1)}$,
using the mapping to a hard-sphere lattice gas described in subsection
\ref{sec:vc}. Using the replica method, a grand-canonical approach
was taken, including a chemical potential $\mu$ controlling the
number of hard spheres. In the limit $\mu\to\infty$, the system tends
to the closest packings, or, equivalently, to the minimal vertex
covers. Assuming the validity of replica symmetry, it was found that 
minimal VCs contain a fraction of vertices given by
\begin{eqnarray}
  \label{eq:minvc}
  x_c(c_0) &=& \lim_{N\to\infty} N^{-1} \min_{\mbox{VC }U} |U|
  \nonumber\\
  &=& 1- \frac{W(c_0)^2+2W(c_0)}{2c_0} 
\end{eqnarray}
with $W(c_0)$ being the real branch of the Lambert-W function defined as
the inverse of $We^W=c_0$. It was also shown, that replica symmetry is
locally stable for $c_0\leq e$, whereas it is unstable for larger
average connectivities, leading to broken replica symmetry. The
correctness of (\ref{eq:minvc}) was recently shown to be exact in
\cite{BaGo2} based on a leaf-removal algorithm, which, as mentioned
above, can be understood as a variant of heuristic-VC.

An interesting insight into the structural properties of minimal
vertex covers was given by identifying a covered and an uncovered {\it
backbone}. The first one is defined as the set of all vertices which
are covered in all minimal VCs, the uncovered backbone unifies all
vertices being {\it uncov} in all minimal VCs. Denoting their relative
sizes by $b_{cov/uncov}(c_0)$, the replica symmetric analysis leads to
\begin{eqnarray}
  \label{eq:bb}
  b_{cov}(c_0) &=&  1- \frac{W(c_0)^2+W(c_0)}{c_0}\nonumber\\
  b_{uncov}(c_0) &=& \frac{W(c_0)}{c_0}\ .
\end{eqnarray}
The remaining $NW(c_0)^2/c$ vertices belong neither to the covered nor
the uncovered backbone, they change the covering state from one
minimal VC to the next. A strong correlation between degree
distribution and backbone was observed: Vertices of small degree tend
to be more frequently in the uncovered backbone, whereas vertices of
high degree can be found more likely in the covered backbone. As
mentioned in subsection \ref{sec:algo}, this can be exploited in the
heuristic algorithm by adapting the selection weights. 


\section{Rate equations for the degree
  distribution}\label{sec:dynamics}

\subsection{Graph reduction dynamics}\label{sec:graphdyn}

We assume that the input to the algorithm heuristic-VC is a random
graph $G=(V,E)$ with $N$ vertices and degree distribution $p_d$, and
we concentrate on the graph reduction process for a moment. The size
of the constructed vertex cover will be calculated in subsection
\ref{sec:vcsize}.

In every algorithmic step, a vertex is selected with weight $w_d$
depending only on its current degree $d$. Then, this vertex and all
its nearest neighbours, 2nd nearest neighbours,... $k$th nearest
neighbours are removed from the graph. The edges incident to these
vertices are removed, too. Following this procedure, a smaller graph
is defined, and the algorithm is iterated.  This graph reduction
process is Markovian, because the action of each algorithmic steps
depends only on the properties of the current reduced graph, more
precisely on the current vertex degrees and neighbourhood relations.

\begin{figure}[htb]
\begin{center}
\includegraphics[width=15cm]{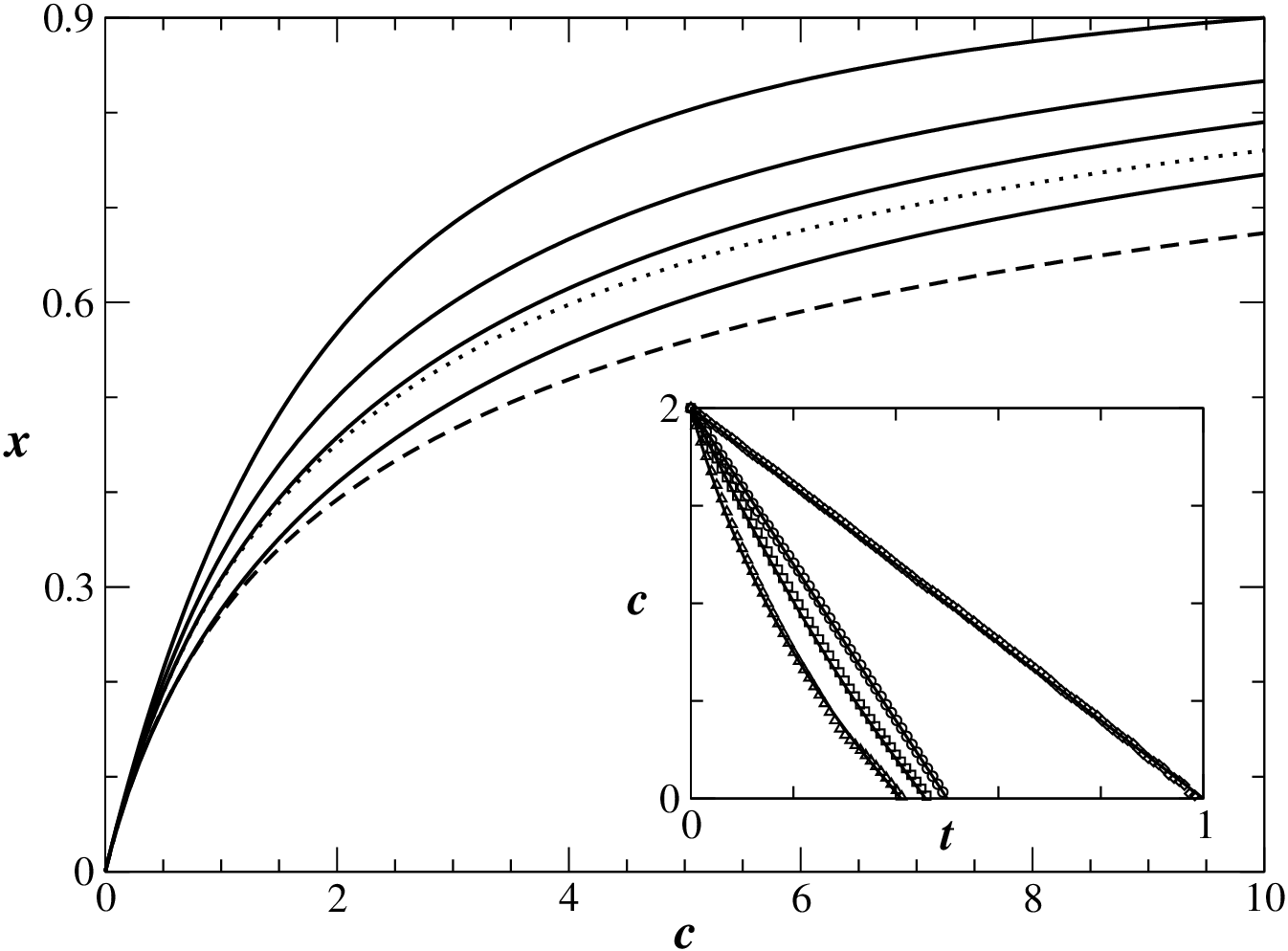}
\end{center}
\caption{Final size $x_f^{(0,\alpha)}(c)$ of the vertex covers
  constructed by the depth-zero heuristic with $\alpha=0,1,2,6$ (full
lines from top to bottom). For a comparison, the replica symmetric
$x_c(c)$ (dashed line, exact for $c<e$) and the results
$x_f^{(1,0)}(c)$ of Gazmuri's depth-one algorithm (dotted line) are
added. The inset shows the time dependent average vertex degree $c(t)$
for the same values of $\alpha$ (full lines from top to bottom)
together with numerical data for a single random graph with $N=3\cdot
10^4$. This illustrates the quality of taking the average trajectory
($\alpha=0,1$), as well as the quality of the binomial approximation
($\alpha=2,6$).}
\label{fig:depthzero}
\end{figure}

Let us further assume, that we start at (algorithmic) time $t=0$, and
every iteration step is counted as $\Delta t$. The dynamics will be
described by rate equations for the vertex-degree distribution $p_d(t)$,
or the number of vertices $N_d(t)=p_d(t)N(t)$ of degree $d$,
where $N(t)$ denotes the remaining vertex number at time $t$. Their
dynamics can be decomposed into the following elementary processes
(where $\la\cdot\ra_t = \sum_{d=0}^\infty(\cdot) p_d(t)$ denotes the
average over the current degree distribution $p_d(t)$):
\begin{itemize}
\item {\it Removal of the central vertex}: A central vertex of degree $d$
  is selected with weight $w_d$, i.e. with probability $w_d p_d(t)/\la
  w_d\ra_t$. $N_d(t)$ is thus reduced by one with this probability.
\item {\it Removal of the 1st, 2nd,..., $k$th nearest neighbours}:
  According to the last item, the central vertex has on average 
  $\la w_d p_d(t) \ra_t/\la w_d\ra_t$ neighbours. As the degrees of
  neighbouring sites are uncorrelated in random graphs, each of these
  has degree $d$ with independent probabilities $p_d^{(1)}(t)=d
  p_d(t)/ \la d \ra_t$. Random graphs are locally tree-like, $d-1$ of
  the $d$ edges of a 1st neighbour lead to 2nd nearest
  neighbours, i.e. the average number of 2nd neighbours equals
  $\frac{\la dw_d \ra_t}{\la w_d \ra_t} \frac{\la d(d-1) \ra_t}{\la d
    \ra_t}$. This argument can easily extended to 3rd neighbours etc.
\item {\it Update of the connectivity of $(k+1)$st neighbours}: The
  edges connecting $k$th and $(k+1)$st neighbours are removed from the
  graph, too. The degree of every $(k+1)$st neighbour is thus reduced
  by one.
\end{itemize}
These processes are combined to evolution equations for the {\it
  expected numbers} $N_d(t)$ of vertices with degree $d$ at time $t$:  
\begin{eqnarray}
  \label{eq:rate}
  N_d(t+\Delta t) &=& N_d(t) - \frac{w_d p_d(t)}{\la w_d \ra_t}
  - \frac{\la d w_d \ra_t}{\la w_d \ra_t} \sum_{m=1}^{k} \left( 
    \frac{\la d (d-1) \ra_t}{\la d \ra_t}\right)^{m-1}   
    \frac{d p_d(t)}{\la d \ra_t}\nonumber\\
  && + \frac{\la d w_d \ra_t}{\la w_d \ra_t}
  \left( \frac{\la d (d-1) \ra_t}{\la d \ra_t}\right)^k  
  \frac{(d+1) p_{d+1}(t) - d p_d(t) }{\la d \ra_t} 
\end{eqnarray}
The first line describes the deletion of vertices, the second the
update of the degrees of all $(k+1)$st neighbours. These equations
are valid for the average trajectory, which is, however, followed
with probability approaching 1 for $N=N(t=0)\to\infty$. Macroscopic
deviations appear only with exponentially small probability and are
thus important for small $N$ only. The quality of using the average
trajectory is demonstrated in the inset of fig. \ref{fig:depthzero}.
There the trajectory of a single graph with $N=3\cdot 10^4$ vertices 
is found to excellently follow the analytical prediction.

Using equations (\ref{eq:rate}), we can calculate also the evolution 
of the total numbers of remaining vertices, $N(t)=\sum_d N_d(t)$,
and edges, $M(t)=\frac12 \sum_d d N_d(t)$:
\begin{eqnarray}
  \label{eq:NM}
  N(t+\Delta t) &=& N(t) - 1 - \frac{\la d w_d \ra_t}{\la w_d \ra_t} 
  \sum_{m=1}^{k}\left(\frac{\la d (d-1) \ra_t}{\la d \ra_t}\right)^{m-1}   
    \nonumber\\
 M(t+\Delta t) &=& M(t) - \frac12  \frac{\la d w_d \ra_t}{\la w_d \ra_t}
 \left(1+\frac{\la d^2 \ra_t}{\la d \ra_t} \sum_{m=1}^{k+1} \left[ 
    \frac{\la d (d-1) \ra_t}{\la d \ra_t}\right]^{m-1} 
   - \left[ 
    \frac{\la d (d-1) \ra_t}{\la d \ra_t}\right]^{k+1} \right)\ .
\end{eqnarray}
As we are mainly interested in the behaviour of large graphs, $N\gg
1$, we may change to intensive quantities by writing $N(t)=n(t)N,\quad
N_d(t)=p_d(t)n(t)N$. Setting further $\Delta t = \frac 1N$, and
replacing differences by derivatives in the thermodynamic limit, we
find 
\begin{eqnarray}
  \label{eq:dgl}
  \dot n(t) &=&  - 1 - \frac{\la d w_d \ra_t}{\la w_d \ra_t} 
  \sum_{m=1}^{k}\left(\frac{\la d (d-1) \ra_t}{\la d \ra_t}\right)^{m-1}   
    \nonumber\\
  \dot n(t)p_d(t) + n(t) \dot p_d(t) &=& 
    - \frac{w_d p_d(t)}{\la w_d \ra_t}
    - \frac{\la d w_d \ra_t}{\la w_d \ra_t} \sum_{m=1}^{k} \left( 
    \frac{\la d (d-1) \ra_t}{\la d \ra_t}\right)^{m-1}   
    \frac{d p_d(t)}{\la d \ra_t}\nonumber\\
  && + \frac{\la d w_d \ra_t}{\la w_d \ra_t}
  \left( \frac{\la d (d-1) \ra_t}{\la d \ra_t}\right)^k  
  \frac{(d+1) p_{d+1}(t) - d p_d(t) }{\la d \ra_t} \ .
\end{eqnarray}
The graph reduction process is thus described by an infinite set of
non-linear differential equations, where the non-linearity enters
only through the time-dependent averages $\la\cdot\ra_t$. As we were
starting with an ordinary random graph, these equations have to be
solved under the initial condition
\begin{eqnarray}
  \label{eq:ic}
  n(0) &=& 1 \nonumber\\
  p_d(0) &=& e^{-c_0} \frac{ c_0^d}{d!} 
\end{eqnarray}
where $c_0$ equals the initial average vertex degree.

In section \ref{sec:nonlinear} we also need the dynamical
equations for the moments $\la d^n \ra_t$ of $p_d(t)$. Multiplying the
second of eqs. (\ref{eq:dgl}) with $d^n$ and summing over all degrees
yields
\begin{eqnarray}
  \label{eq:dglmoment}
  n(t) \frac d{dt} \la d^n \ra_t &=& \frac{ \la w_d \ra_t \la d^n
    \ra_t - \la d^n w_d \ra_t}{\la w_d \ra_t} + \frac{\la d w_d
    \ra_t}{\la w_d \ra_t} \frac{\la d \ra_t\la d^n \ra_t-\la d^{n+1}
    \ra_t}{\la d \ra_t} \sum_{m=1}^k \left( \frac{\la d(d-1) \ra_t}{
      \la d \ra_t} \right)^{m-1} \nonumber\\
 && + \frac{\la d w_d
    \ra_t}{\la w_d \ra_t} \frac{\la d(d-1)^n \ra_t-\la d^{n+1}
    \ra_t}{\la d \ra_t} \left( \frac{\la d(d-1) \ra_t}{
      \la d \ra_t} \right)^k
\end{eqnarray}
Please note that these equations do not contain any finite and closed
subset of equations, because the evolution of any moment depends also 
on higher moments.

A similar approach was chosen in \cite{PiSpWo} to analyze an algorithm
constructing the maximal sub-graph having minimal degree $K$,
i.e. the so-called $K$-core. Pittel et al. rigorously derived and
solved a closed set of equations for $N_0(t),...,N_{K-1}(t)$ and
$M(t)$. An analogous reduction to a finite number of equations will be
constructed in section \ref{sec:leaf} for the generalized leaf-removal
algorithm, but it cannot be achieved for the general case.

Our approach resembles also the rate equation approach used in the
area of growing networks \cite{AlBa,KrRe,DoMe}. Note, however, that
there the evolution of the number $N_d(t)$ of vertices of degree $d$
depends only on the vertices having smaller degree, because edges are
always added and never deleted. So, in principle, the evolution
equations can be solved by calculating first $N_0(t)$, then $N_1(t)$
and so on.  In the graph reduction process the problem becomes more
complicated since the evolution of $N_d$ depends also on $N_{d+1}$,
for arbitrary $d$, but there is no maximal degree $d$ in the
Poissonian initial condition.

\subsection{The cardinality of constructed vertex covers}
\label{sec:vcsize}

Before trying to solve equations (\ref{eq:dgl},\ref{eq:ic}) for
specific choices of $w_d$ and $k$, we will give general expressions
for the number $X(t)$ of vertices which are covered by the algorithm.

For the locally tree-like case of random graphs, the $k$th neighbours
of the selected vertex are covered, the $(k-1)$st are uncovered etc.
So the covering state of the central vertex depends on the depth $k$:
For even $k$, it will be covered (if $d\neq 0$), for odd $k$, it will
be uncovered.  We therefore consider these two cases independently.

\subsubsection{Odd depth $k$}
For odd $k$, the central vertex is almost always uncovered. Denoting
the expected number of covered vertices at time t with $X(t)=x(t) N$, 
we thus find
\begin{equation}
  \label{eq:Xodd}
  X(t+\Delta t) = X(t) + \frac{\la d w_d \ra_t}{\la w_d \ra_t} 
  \sum_{m=0}^{(k-1)/2}\left(\frac{\la d (d-1) \ra_t}{\la d
      \ra_t}\right)^{2m} \ ,
\end{equation}
cf. the first of equations (\ref{eq:NM}). Going again to the limit
of large graphs, $N\to\infty$, this can be written as a differential
equation for $x(t)$:
\begin{equation}
  \label{eq:xodd}
  \dot x(t) = \frac{\la d w_d \ra_t}{\la w_d \ra_t} 
  \sum_{m=0}^{(k-1)/2}\left(\frac{\la d (d-1) \ra_t}{\la d
      \ra_t}\right)^{2m}\ .
\end{equation}
Once we know the solution of the graph dynamical equation
(\ref{eq:dgl}), we can calculate the time $t_f$ where all edges are
covered, $\la d \ra_{t_f}=0$, and integrate the last equation over
the time interval $[0,t_f]$. As all removed edges were covered by
our algorithm, we thus have constructed a vertex cover of
relative size $x(t_f)$. As the described average trajectory is
followed with probability one for $N\to \infty$, this $x(t_f)$ gives
an upper bound for the true minimal vertex cover size of the random
graph under consideration.

\subsubsection{Even depth $k$}
For even $k$, the central vertex is covered in general. Only if it is
disconnected, i.e. if its degree equals zero, it is set to {\it
uncov}, see the algorithm heuristic-VC. The last case happens with
probability $\frac {w_0 p_0(t)}{\la w_d \ra_t}$. We thus conclude for
$X(t)$
\begin{equation}
  \label{eq:Xeven}
  X(t+\Delta t) = X(t) + 1 - \frac {w_0 p_0(t)}{\la w_d \ra_t}
  +\frac{\la d w_d \ra_t}{\la w_d \ra_t} 
  \sum_{m=0}^{k/2-1}\left(\frac{\la d (d-1) \ra_t}{\la d
      \ra_t}\right)^{2m+1} \ ,
\end{equation}
or, in the limit $N\to\infty$,
\begin{equation}
  \label{eq:xeven}
  \dot x(t) = + 1 - \frac {w_0 p_0(t)}{\la w_d \ra_t}
  +\frac{\la d w_d \ra_t}{\la w_d \ra_t} 
  \sum_{m=0}^{k/2-1}\left(\frac{\la d (d-1) \ra_t}{\la d
      \ra_t}\right)^{2m+1}\ .
\end{equation}
This equation can be integrated, once we know the solution of
equations (\ref{eq:dgl}), and an upper bound to $x_c(c_0)$ can be read 
off.


\section{The solvable case of linear selection weights $w_d$}
\label{sec:linear}

The problem in solving differential equations (\ref{eq:dgl}) with
initial conditions (\ref{eq:ic}) is, that the Poissonian shape
of the degree distribution is, in general, not conserved under the
dynamics. In such cases one has to keep track of all the individual
probabilities $p_d(t)$ for each possible degree $d$. As the dynamics
of $p_d(t)$ depends on $p_{d+1}(t)$ for all $d$, and $d$ is not
bound from above for the Poissonian initial condition, it is not
obvious how to construct a finite and closed subset of equations which
can be solved separately, opening the door to the solution for all
$p_d$.

There exist, however, some cases where the Poissonian shape of the
degree distribution is conserved, as can be shown explicitly by
plugging a Poissonian Ansatz 
\begin{equation}
  \label{eq:poissonansatz}
  p_d(t) = e^{-c(t)} \frac{c(t)^d}{d!}
\end{equation}
into (\ref{eq:dgl}) and verifying, that the same equation for the 
average vertex degree $c(t)$ is reproduced for 
arbitrary $d$. The most general case for this behaviour is found for
linear selection weights
\begin{equation}
  \label{eq:linear}
  w_d = A\cdot d+B
\end{equation}
where $A,B$ are arbitrary non-negative real numbers. In this case, the
graph can be totally specified by calculating $n(t)=\frac{N(t)}N$ and 
$c(t)=\frac{2M(t)}{N(t)}$. Their evolution can be read off
from equations (\ref{eq:NM}), where the averages $\la\cdot\ra_t$ can
be expressed via $c(t)$:
\begin{eqnarray}
  \label{eq:avs}
  \la d \ra_t &=& c(t) \nonumber\\
  \la d(d-1) \ra_t &=& c(t)^2 \nonumber\\
  \la w_d \ra_t &=& A c(t)+B \nonumber\\
  \la dw_d \ra_t &=& Ac(t)^2+(A+B)c(t)
\end{eqnarray}
In the limit $N\to\infty$ the graph reduction dynamics is thus
completely determined by the differential equations
\begin{eqnarray}
  \label{eq:dgllin}
  \dot n(t) &=&  - 1 - \frac{Ac(t)^2+(A+B)c(t)}{ A c(t)+B} 
  \sum_{m=0}^{k-1}c(t)^m   
    \nonumber\\
  \dot n(t) c(t) + n(t) \dot c(t) &=&  
      -2\frac{Ac(t)^2+(A+B)c(t)}{ A c(t)+B} \sum_{m=0}^{k}c(t)^m     
\end{eqnarray}
Eliminating $\dot n(t)$ from the second equation, we end up with
\begin{equation}
  \label{eq:cdgl}
  n(t)\dot c(t) = -\frac{Ac(t)^2+(2A+B)c(t)}{ A c(t)+B} 
  -\frac{Ac(t)^2+(A+B)c(t)}{ A c(t)+B} \sum_{m=1}^{k}c(t)^m\ .
\end{equation}
These equations have to be solved under the initial conditions
$n(t=0)=1$ and $c(t=0)=c_0$

\subsection{Constant selection weights: $A=0,\quad B=1$}
\label{sec:const}

Equations (\ref{eq:dgllin},\ref{eq:cdgl}) simplify further if we
restrict it for a while to constant selection weights $w_d\equiv 1$,
i.e. $A=0,\quad B=1$. There we find
\begin{eqnarray}
  \label{eq:dglconst}
  \dot n(t) &=&  - \sum_{m=0}^{k}c(t)^m   
    \nonumber\\
  n(t) \dot c(t) &=&  c(t) \dot n(t)\ .
\end{eqnarray}
In the second line, the equation for $\dot n(t)$ was already used 
to eliminate the complicated sum of powers of $c(t)$. Using the 
initial conditions $n(t=0)=1$ and $c(t=0)=c_0$, the second line
results in
\begin{equation}
  \label{eq:nc}
  c(t) = c_0 n(t)\ ,
\end{equation}
and one of the two functions can be eliminated from the first of
equations (\ref{eq:dglconst}). We consequently find
\begin{equation}
  \label{eq:cdot}
  \dot c(t) = - c_0\sum_{m=0}^{k}c(t)^m   
\end{equation}
which is solved implicitly by
\begin{equation}
  \label{eq:ct}
  t = \frac{1}{c_0} \int_{c(t)}^{c_0}\frac{d\tilde
    c}{\sum_{m=0}^{k}\tilde c^m}\ .   
\end{equation}
The algorithm stops when all edges are covered, i.e. for $c(t_f)=0$.
This final time $t_f$ corresponds, in the original algorithmic
language, to $t_f N$ iterations of heuristic-VC, and is given by
\begin{equation}
  \label{eq:tf}
  t_f = \frac{1}{c_0} \int_{0}^{c_0}\frac{d\tilde
    c}{\sum_{m=0}^{k}\tilde c^m}\ .   
\end{equation}

These results can be used in order to determine the relative size
$x_f(c_0)= x(t_f)$ using the results of section \ref{sec:vcsize}.
There we observed a difference between even and odd
values $k$ of the depth of heuristic-VC according to the fact that
the central vertex in one case is almost always covered, in the other
case uncovered. We therefore continue discussing these cases
separately, starting with $k=0$, going then to arbitrary odd $k$,
and discussing general even $k$ at the end of this section.

\subsubsection{The simplest algorithm}

The simplest possible algorithm has depth $k=0$: In every algorithmic
step an arbitrary vertex is chosen and covered if its degree is
non-zero, uncovered else. The vertex and all incident edges are
removed from the graph. This simple algorithm was already analysed in
\cite{WeHa4} as the heuristic underlying a complete backtracking
algorithm. The results given there can be easily reproduced, for $k=0$
the integration in eq. (\ref{eq:ct}) can be trivially carried out.
We find a simple linear decrease of the average vertex degree,
\begin{equation}
  \label{eq:ct_w1_k0}
  c(t) = (1-t) c_0\ ,
\end{equation}
and the final time becomes $t_f=1$ as one vertex is removed in every
algorithmic step. The resulting size of the constructed vertex cover
follows easily by integrating equation (\ref{eq:xeven}), with
$p_0(t)=e^{-c(t)}$:
\begin{equation}
  \label{eq:x_w1_k0}
  x_f^{(0,0)}(c_0) = 1- \frac{1-e^{-c_0}}{c_0}\ ,
\end{equation}
cf. \cite{WeHa4}. This gives the very first and simplest upper bound
on the true size of minimal vertex covers which is, however, not very
good: For, e.g., $c_0=2$, we find $x_f(2)=0.5677$ compared to the true
value $x_c(c)=0.3919$, cf. eq. (\ref{eq:minvc}). Also the asymptotic
behaviour for large average vertex degrees, $x_f(c_0>>1)\simeq
1-\frac1{c_0}$, does not meet the exact asymptotic behaviour, which
was evaluated by Frieze \cite{Fr} to be $x_c(c_0>>1)\simeq
1-\frac{\ln c_0}{c_0}$ to leading order.

\subsubsection{Gazmuri's algorithm and odd depths}

This asymptotics is found to hold also for a slightly more
complicated case: It is valid for all non-zero values of the depth,
i.e. for $k>0$. Let us, for simplicity, start with arbitrary odd
values of the depth $k$ of the algorithm. For $k=1$, the algorithm is
equivalent to the one proposed by Gazmuri \cite{Ga}: In every time
step, an arbitrary vertex is chosen and set to \textit{uncov}, all
its neighbours are covered, and the whole cluster including all
incident edges are removed from the graph. Gazmuri has already
characterized the performance of this algorithm, using a different
technique.

Please remember that, according to equation (\ref{eq:cdot}), the time
dependency of the mean vertex degree is given by
\begin{equation}
  \label{eq:cdot1}
  \dot c(t) = - c_0\sum_{m=0}^{k}c(t)^m   \ .
\end{equation}
This can be used to solve equation (\ref{eq:xodd}) for the
evolution of the number of covered vertices, which, using the
Poissonian distribution (\ref{eq:poissonansatz}), and integrating
over $t$ reads 
\begin{equation}
  \label{eq:xt}
  x(t) = \int_0^t dt' \sum_{m=0}^{(k-1)/2} c(t')^{2m+1}\ .
\end{equation}
Changing variables from $t$ to $c$, and plugging in equation
(\ref{eq:cdot1}), we find
\begin{eqnarray}
  \label{eq:xc}
  x(c) &=& \frac1{c_0} \int_c^{c_0} d\tilde c \frac
  {\sum_{m=0}^{(k-1)/2} \tilde c^{2m+1} }{\sum_{n=0}^{k} \tilde c^n
    }\nonumber\\ 
 &=&\frac1{c_0} \int_c^{c_0} d\tilde c \frac {\tilde c}{1+\tilde c}
 \nonumber\\ 
 &=& \frac{c_0-c}{c_0} - \frac1{c_0} \ln \frac{1+c_0}{1+c} \ .
\end{eqnarray}
This expression gives the $x$-$c$-trajectory which is interestingly
independent on the depth $k$ (as long as $k$ is odd). The only
difference is given by the time dependencies $x(t)=x(c(t))$ as
$c(t)$ is $k$-dependent, cf. equation (\ref{eq:ct}). 

The graph is completely covered when $c(t)$ reaches zero. The
cardinality of this vertex cover is almost surely given by
$x_f(c_0)=x(c=0)$, i.e.
\begin{equation}
  \label{eq:xf1}
  x_f^{(k,0)}(c_0) = 1 - \frac{\ln(1+c_0)}{c_0}
\end{equation}
independently on the (odd) depth $k$ of the algorithm. For $k=1$,
Gazmuri's bound is thus reproduced, and the leading order of the
behaviour for large initial connectivities $c_0$ is correctly
found. The algorithm is, however, less successful for small and
intermediate connectivities, as we will see in the following
sections. 

\subsubsection{The case of even depth $k\geq 2$}

The case of even depth leads to more complicated expressions, which
cannot be evaluated explicitly. The main problem is induced by the
$p_0(t)$-contribution in the evolution of $x(t)$, as given in equation
(\ref{eq:xeven}). After having applied the Poissonian ansatz
(\ref{eq:poissonansatz}), the latter reads
\begin{equation}
  \label{eq:xteven}
  x(t) = \int_0^t dt' \left( \sum_{m=0}^{k/2} c(t')^{2m} - 
  e^{-c(t)} \right) \ ,
\end{equation}
hence we find for the relative vertex-cover size, and arbitrary even
depth $k$
\begin{equation}
  \label{eq:xf2}
  x_f^{(k,0)}(c_0) = \frac1{c_0} \int_0^{c_0} d\tilde c \frac
  {\sum_{m=0}^{k/2} \tilde c^{2m}-e^{-\tilde c} }{\sum_{n=0}^{k} 
  \tilde c^n }\ .
\end{equation}
This integral cannot be evaluated explicitly for arbitrary even
$k$. We can, however, extract the asymptotic behaviour for $c_0\gg
1$. In this limit, the terms of $O(e^{-\tilde c})$ and $O(\tilde c^0)$
can be neglected in both the numerator and the denominator in the last
integral. The corrections from the integration interval
$(0,O((c_0)^0))$ are of order 1 and thus suppressed by the prefactor
$\frac1{c_0}$, compared to the leading terms
\begin{equation}
  \label{eq:xf2asympt}
  x_f^{(k,0)}(c_0\gg 1) \simeq 1 - \frac{\ln(1+c_0)}{c_0}
\end{equation}
which coincide with the case of odd depth.

\subsection{Linear selection weights: $A=1,\quad B=0$}
\label{sec:lin}

After having discussed in great detail algorithms with $w_d=const.$,
i.e. simple algorithms selecting central vertices completely at
random, without regarding its degree, we now turn to the case of
linear $w_d$. In this case, as already mentioned at the beginning of
this section, the degree distribution still stays Poissonian. This can
be understood intuitively in the following way: The case $w_d = d$ is
equivalent to choosing an arbitrary edge with constant probability,
and selecting one of the end-vertices. According to section
\ref{sec:randomgraph}, the chosen vertex will have degree $d$ with
probability $p_d^{(1)}\propto dp_d$.  The more general case
$w_d=Ad+B$, cf. eq. (\ref{eq:linear}), corresponds to mixing this
selection procedure (weight $A$) with the uniform selection of
vertices (weight $B$).

Fixing $A$ to a non-zero value is sensible only for \textit{even
depth} values $k$ since these correspond to covering the central
vertex of the cluster of radius $k$. As mentioned above, the vertex
degree is correlated to the covering state in small vertex covers, so
it does not make sense to preferentially select vertices of high
degree and to uncover it subsequently. The performance of the
algorithm is thus only improved for even $k$. The strongest
improvement is obtained for $w_d=d$, i.e. for $B=0$.

Whereas eqs. (\ref{eq:dgllin},\ref{eq:cdgl}) lead to complicated
coupled non-linear differential equations for $c(t)$ and $n(t)$,
which have to be solved numerically for general even $k$, the 
case $k=0$ becomes very simple. There, the equations read
\begin{eqnarray}
  \label{eq:wdk0}
  \dot n(t) &=& -1 \nonumber\\
  n(t)\dot c(t) &=& -c(t) - 2 \nonumber\\
  \dot x(t) &=& 1
\end{eqnarray}
and are solved by
\begin{eqnarray}
  \label{eq:wdk0sol}
  n(t) &=& 1-t \nonumber\\
  c(t) &=& c_0 - (2+c_0) t \nonumber\\
  x(t) &=& t
\end{eqnarray}
The graph is covered for $c(t_f)=0$, which leads to vertex covers 
of size
\begin{equation}
  \label{eq:wdk0vc}
  x_f^{(0,1)}(c_0) = 1-\frac2{2+c_0}\ .
\end{equation}
This size is always smaller than the one found for the very simplest
algorithm ($w_d=1, \ k=0$), but stays worse than Gazmuri's
algorithm ($w_d=1, \ k=1$). Also the asymptotic behaviour is not
correctly reproduced.


\section{An approximation for depth-zero algorithms with non-linear
selection weights}
\label{sec:nonlinear}

If we choose non-linear selection weights $w_d=d^\alpha,\ \alpha\neq
0,1$, the graph reduction dynamics deviates from the unrestricted
ensemble of random graphs, and the degree distribution becomes
non-Poissonian. The new distribution thus cannot be described by the
evolution of its mean value alone, and we have to solve all equations from
(\ref{eq:dgl}) simultaneously. In general, this cannot be achieved
analytically. To approximate the solution numerically, we may cut the
tail of $p_d(t)$, and solve only a finite number of equations. This
works fine for small initial values of $c_0=c(t=0)$ because of the
rapid decrease of the Poissonian distribution. For larger values of
$c_0$ however, the number of remaining equations becomes large,
too. So it would be better to find an approximation of $p_d(t)$
depending only on a small number of parameters. The dynamics of these
parameters can be determined from the lowest moments $\la d^m \ra_t,$ 
with $m$ ranging from 1 to the number of parameters.

This section is dedicated to algorithms of depth $k=0$, but non-linear
selection weights $w_d = d^\alpha,\ \alpha\neq 0,1$. Every algorithmic
step thus removes one vertex from the graph. No vertices of degree 0
are selected ($w_0=0$); all selected vertices thus have to be covered
according to heuristic-VC. We thus trivially have
\begin{eqnarray}
  \label{eq:k0nx}
  n(t) &=& 1-t \nonumber\\
  x(t) &=& t
\end{eqnarray}
and the size of the constructed vertex cover is given by
$x_f^{(0,\alpha)}(c_0)=t_f$ with $c(t_f)=0$. Note that this does not
necessarily imply $n(t_f)=0$ as only vertices of non-vanishing degree
are selected. At time $t_f$, the remaining graph consists of
$(1-t_f)N$ completely disconnected vertices.

As the initial condition $p_d(t=0)$ of the graph reduction dynamics is
given by a Poissonian of mean $c_0$, we are looking for a deformation
of the Poissonian distribution which allows to independently vary mean
and variance, and thus to approximate the true $p_d(t)$. A simple 
possibility is given by the generalized binomial expression
\begin{equation}
  \label{eq:binomial}
  \pi_d(t) = \frac{\Gamma(1+\mu(t)^{-1})}{\Gamma(1+\mu(t)^{-1}-d)
    \Gamma(1+d)}\ \left[1-c(t)\mu(t)\right]^{(\mu(t)^{-1}-d)} 
    \ \left[c(t)\mu(t)\right]^d\ .
\end{equation}
For $\mu(t)\to 0$, this expression approaches a Poissonian
distribution of mean $c(t)$, the initial condition is thus
characterized by $c(t=0)=c_0$ and $\mu(t=0)=0$. Please note, however,
that (\ref{eq:binomial}) does not necessarily describe a probability
distribution, because $\pi_d(t)$ becomes negative for certain $d$ if
$1/\mu(t)$ is not a positive integer. We do not expect this to produce
serious problems for the calculated averages, as long as the absolute
value of these negative $\pi_d(t)$ stays neglectable compared to the
total normalization 1. This is exactly what happens in our case, as we
will see below.

The moments of $\pi_d(t)$ are given by
\begin{equation}
  \label{eq:moment1}
  \Pi^{(n+1)}(t) := \sum_{d=0}^\infty d(d-1)\cdots(d-n) \pi_d(t)
  = c(t)^{n+1} [1-\mu(t)][1-2\mu(t)]\cdots[1-n\mu(t)]\ .
\end{equation}
The product $d(d-1)\cdots(d-n)$ can be expanded into a sum of pure
powers, their mean values are determined by inverting this sum:
\begin{equation}
  \label{eq:moment}
  \tilde \Pi^{(n)}(t) :=\sum_{d=0}^\infty d^n \pi_d(t)
  = \sum_{m=1}^{n} a_m^{(n)} \Pi^{(n)}(t)\ .
\end{equation}
The coefficients are given iteratively by 
$a_m^{(n)}=m a_m^{(n-1)}+a_{m-1}^{(n-1)}$, using the trivial identity
$a_m^{(1)}=\delta_{m,1}$ and the convention $a_0^{(n)}\equiv 0$ for
all $n$.

\begin{figure}[htb]
\begin{center}
\includegraphics[width=15cm]{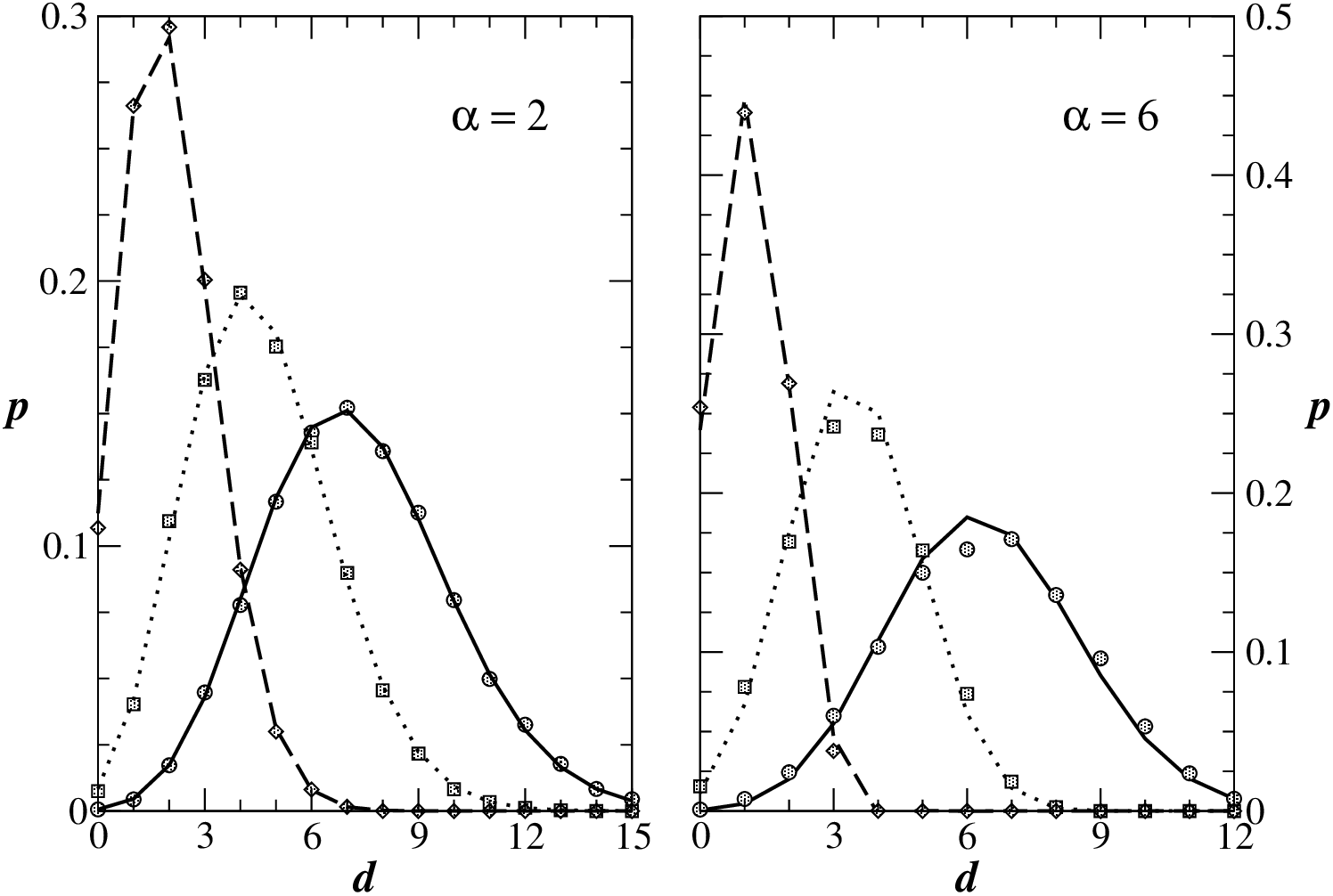}
\end{center}
\caption{(a) Degree distribution $p_d(t)$ for heuristic-VC
  with depth $k=0$, selection weights $w_d=d^2$, and initial condition
  $c_0=10$, for $t=0.2,0.4,0.6$ (full, dotted, dashed lines). The symbols
  are numerical data for a single graph of size $N=3\cdot10^4$ and
  coincide extremely well with the binomial approximation represented by
  the lines (lines are connecting data for integer $d$ and thus are
  guides to the eyes only). The quality of the approximation is
  similarly good for all investigated initial $c_0$.\\
  (b) Same as (a), but with $w_d=d^6$. The coincidence between
  numerical data and binomial approximation is slightly worse than
  in (a), but still very convincing. The quality of the approximation 
  increases with growing initial $c_0$, supporting thus our
  conjecture that the asymptotic behaviour is correctly described by
  the approximation.}
\label{fig:distr}
\end{figure}

Given two non-trivial moments, e.g. $n=1,2$, $c(t)$ and $\mu(t)$ can
be calculated, and all other moments are determined. We can thus
approximate the dynamics of the degree distribution by
considering of the dynamics of the first two moments of $p_d(t)$
only. The exact equations for $\la d \ra_t$ and $\la d(d-1)\ra_t$
follow from eq. (\ref{eq:dglmoment}) by fixing $k=0$ and
$w_d=d^\alpha$,
\begin{eqnarray}
  \label{eq:2moments}
  (1-t) \frac d{dt} \la d \ra_t &=& \la d \ra_t -2 \frac{\la
    d^{\alpha+1} \ra_t}{\la d^\alpha \ra_t} \nonumber\\
 (1-t) \frac d{dt} \la d(d-1) \ra_t &=&\la d(d-1) \ra_t -\frac{\la 
    d^{\alpha+2} \ra_t}{\la d^\alpha \ra_t} + \frac{\la
    d^{\alpha+1} \ra_t}{\la d^\alpha \ra_t} \left( 1-2 \frac{\la
      d(d-1) \ra_t}{\la d \ra_t } \right)\ .
\end{eqnarray}
They contain higher moments of $p_d(t)$, but are closed
approximately by using (\ref{eq:moment1},\ref{eq:moment}):
\begin{eqnarray}
  \label{eq:binomapprox}
  (1-t) \frac d{dt} c(t) &=& c(t) - 2 \frac {\tilde
    \Pi^{(\alpha+1)}(t)}{\tilde \Pi^{(\alpha)}(t)} \nonumber\\
  (1-t) \frac d{dt} \left( c(t)^2 [1-\mu(t)] \right) &=& 
    c(t)^2 [1-\mu(t)] - \frac {\tilde \Pi^{(\alpha+2)}(t)}{\tilde 
    \Pi^{(\alpha)}(t)} + \frac {\tilde \Pi^{(\alpha+1)}(t)}{\tilde 
    \Pi^{(\alpha)}(t)}\left( 1-2 c(t) [1-\mu(t)] \right)
\end{eqnarray}
These two equations can be easily converted to ordinary differential
equations $\dot c(t) = F_c ( c(t), \mu(t), t)$ and $\dot \mu(t) =
F_\mu ( c(t), \mu(t), t)$ which are, however, not analytically
solvable for general values of $\alpha$. They can instead be solved
efficiently using numerical standard techniques. The results are
displayed in fig. \ref{fig:depthzero} together with numerical
simulations obtain for large graphs. We find that the binomial
approximation works extremely well for small values of $\alpha$,
systematically growing deviations appear for larger $\alpha$,
cf. fig. \ref{fig:distr}. We also find that heuristic-VC is able to
approximate the true minimal vertex-cover size up to a few percent,
performing better for larger $\alpha$. We observe, however, that even
the vertex covers constructed for very large $\alpha$ remain
suboptimal, i.e. extensively larger than the minimal VCs.

The behaviour of eqs. (\ref{eq:binomapprox}) for large $c(t=0)=c_0$
can be extracted analytically by expanding the equations for $\dot
c(t)$ ($\dot \mu(t)$) to $O(1/c)$ ($O(1/c^2)$). $F_\mu$ is a sum of terms of
$O(\mu^2)$, $O(\mu/c)$ and $O(1/c^2)$, and $\mu(t)$ stays of
$O(1/c^2)$ due to its initial condition $\mu(0)=0$. We thus find
$(1-t)\dot c = -c-2\alpha-O(1/c)$ which is solved to leading orders
by $c(t)\simeq c_0 - t(c_0+2\alpha)$. From the vanishing of these
leading orders we can read of the dominant contributions to the
constructed vertex covers, which we conjecture to be exact also
for the true dynamics:
\begin{equation}
  \label{eq:largec}
  x_f^{(0,\alpha)}(c_0\gg 1) \simeq 1 - \frac{2\alpha}{c_0+2\alpha}\ .  
\end{equation}
For larger $\alpha$, the asymptotics becomes better and better, but
still does not reach the correct behaviour $x_c \simeq 1 -
\ln(c_0)/c_0$ of the minimal VCs. Note that, for sufficiently large
$c_0$, Gazmuri's depth-one algorithm outperforms the depth-zero
algorithm for arbitrary $\alpha$. We expect, however, that the
correct asymptotic behaviour is reached by exponential selection 
weights $w_d$, and hence in particular for the heuristic where always 
a vertex of maximal degree is chosen.


\section{Generalizing the leaf removal procedure: Depth-one
  algorithms} 
\label{sec:leaf}

The best performance can be achieved using a generalization of the
leaf-removal algorithm (LR) proposed by Bauer and Golinelli \cite{BaGo1}.
Their algorithm is based on the following observation: Consider a
vertex of degree one, i.e. a vertex which has a single neighbour. One
of these two vertices has to be covered in order to cover also the
connecting edge. It is obviously better to cover the neighbour than
the vertex of degree one. Iterating this procedure, there are two
possible final situations:\\
(i) All edges are covered, and the constructed vertex cover is a
minimal one.\\
(ii) There are uncovered edges, but no vertices of current degree one 
are left. The algorithm stops without having constructed a vertex
cover.\\
This algorithm can be understood as a special case of heuristic-VC
with depth $k=1$ and $w_d = \delta_{d,1}$. Following a completely
different route, Bauer and Golinelli have found a
surprising result: For random graphs of average degree $c_0<e$, their
algorithm is able to cover almost all edges, and the predicted minimal
vertex cover size coincides with the replica symmetric one found in
\cite{WeHa1}. For larger average degrees, $c_0>e$, case (ii) is valid,
and a finite fraction of edges remains uncovered. We re-derive this
result below using our dynamical rate equations.

In order to construct a small vertex cover also for
higher-connectivity graphs, the algorithm has to be modified to
$w_d>0$ for all $d>0$. The case $w_d = A \delta_{d,1} + 1$, with
$A>0$, will therefore be analysed in sec. \ref{sec:leafgeneral}. For
$A\gg 1$, this algorithm performs nearly as well as pure
leaf-removal for small average vertex degrees. For large $c_0$ it
finally becomes more and more similar to Gazmuri's algorithm, being
still better for any finite $c_0$. We will call this algorithm {\it
  generalized leaf removal} (GLR).

The best performance is, of course, obtained for $w_d = e^{-\alpha
d}$, with $\alpha\to\infty$. There always a vertex of minimal degree
is selected and uncovered, all neighbours are covered. As long as the
fraction $p_1(t)$ of vertices of degree one is non-zero, the algorithm
is equivalent to the leaf-removal procedure. This is valid in
particular also for $c_0<e$, where (almost) minimal VCs are
constructed. The analysis of this algorithm goes, however, beyond the
analysis presented here.

\subsection{Leaf removal}
\label{sec:bago}

Let us first concentrate on the leaf-removal algorithm (LR) in its
original version, i.e. on depth $k=1$ and selection weights
$w_d=\delta_{d,1}$. In every algorithmic step, exactly two vertices
are removed from the graph $G$, and exactly one vertex is covered.
We therefore conclude
\begin{eqnarray}
  \label{eq:nx}
  n(t) &=& 1-2t \nonumber\\
  x(t) &=& t
\end{eqnarray}
The degree distribution follows, according to (\ref{eq:dgl}), 
from the dynamical equations
\begin{equation}
  \label{eq:LRdgl}
  (1-2t) \dot p_d (t) = 2 p_d(t) - \delta_{d,1} - \frac{\la d^2
    \ra_t}{ \la d \ra_t^2} dp_d(t) + \frac{\la d(d-1) \ra_t}{ 
    \la d \ra_t^2} (d+1) p_{d+1}(t)\ .
\end{equation}
Vertices of degree $d>1$ are only touched if they are first or second
neighbours of a vertex of degree one, in which cases they are either
covered and removed, or their vertex degree is reduced by one. The
degrees of neighbouring vertices are statistically independent, we
thus expect $p_d(t)$ to keep its Poissonian shape for all $d>1$. In
fact, the ansatz
\begin{eqnarray}
  \label{eq:LRansatz}
  p_d(t) = \gamma(t) e^{-\kappa(t)} \frac{ \kappa(t)^d}{d!}
\qquad \forall d > 1
\end{eqnarray}
together with the global normalization
\begin{equation}
  \label{eq:norm}
  1 = p_0(t) + p_1(t) + \gamma(t) \left( 1- e^{-\kappa(t)}
  [1-\kappa(t)] \right)
\end{equation}
can be plugged into eqs. (\ref{eq:LRdgl}) and leads to uniquely
determined equations for $\gamma(t),\ \kappa(t)$ and $p_1(t)$. The
latter is, for technical reasons, replaced by an equation for $c(t)$.
Using 
\begin{eqnarray}
  \label{eq:LRmoments}
  \la d \ra_t &=& c(t) \nonumber\\
  &=& p_1(t) + \gamma(t) \kappa(t) \left[ 1-e^{-\kappa(t)} \right]
  \nonumber\\
  \la d(d-1) \ra_t &=&  \gamma(t) \kappa(t)^2\ ,
\end{eqnarray}
these read
\begin{eqnarray}
  \label{eq:LRreduceddgl}
  (1-2t) \dot c(t) &=& 2 c(t) - 2 \frac{
    \gamma(t)\kappa(t)^2+c(t)}{c(t)} \nonumber\\
  (1-2t) \dot \kappa(t) &=& -\kappa(t) \frac{
    \gamma(t)\kappa(t)^2+c(t)}{c(t)} \nonumber\\
  (1-2t) \dot \gamma(t) &=& \gamma(t) \frac{ 2c(t)-\kappa(t)}{c(t)} 
\ .
\end{eqnarray}
The initial conditions are $c(0)=\kappa(0)=c_0,\ \gamma(0)=1$.
The equation for the average vertex degree $c(t)$ can be removed by
observing 
\begin{equation}
  \label{eq:LRcelim}
  (1-2t) \frac d{dt} \left( \frac{\kappa(t)^2}{c(t)} \right)
  = (1-2t) \frac{\kappa(t)^2}{c(t)} \left( 2 \frac{\dot
      \kappa(t)}{\kappa(t)} -\frac{\dot c(t)}{c(t)} \right)
  = -2 \frac{\kappa(t)^2}{c(t)}
\end{equation}
which is solved by
\begin{equation}
  \label{eq:LRcsol}
  c(t) = \frac{ \kappa(t)^2}{(1-2t)c_0}\ .
\end{equation}
The solution of the two remaining equations is given implicitly by
\begin{eqnarray}
  \label{eq:LRsol}
  t &=& 1 - \frac1{2c_0} \left( [\kappa(t)-W(c_0e^{\kappa(t)})]^2 
        +2W(c_0e^{\kappa(t)}) \right) \nonumber\\
  \gamma(t) &=& \frac{W(c_0e^{\kappa(t)}) }{(1-2t)c_0}
\end{eqnarray}
as can be checked explicitly using eqs. (\ref{eq:LRreduceddgl}). 
$W$ denotes again the Lambert-W function, cf. sec. \ref{sec:minvc}.
The graph is covered if this trajectory reaches $c(t_f)=0$, i.e. for 
$\kappa(t_f)=0$. From the first of eqs. (\ref{eq:LRsol}) we thus find
the final time
\begin{equation}
  \label{eq:LRtf}
   t_f = 1 - \frac{W(c_0)^2 +2W(c_0)}{2c_0}\ .
\end{equation}
This result is only valid, if $p_d(t)\geq 0$ for all $d$ and all
$0<t<t_f$. Using eq. (\ref{eq:LRmoments}), we find
\begin{eqnarray}
  \label{eq:LRp1}
  p_1(t) &=& c(t) - \gamma(t) \kappa(t) \left[ 1-e^{-\kappa(t)}
  \right] \nonumber\\
  &=& \frac{ \kappa(t)}{(1-2t)c_0} \left( \kappa(t) - W(c_0
    e^{\kappa(t)}) [1-e^{-\kappa(t)}]  \right) \nonumber\\
  &=:&  \frac{ \kappa(t)}{(1-2t)c_0} \Phi( \kappa(t) )\ .
\end{eqnarray}
The prefactor of $\Phi(\kappa)$ is non-negative, it is thus sufficient
to investigate $\Phi(\kappa)$ for $\kappa\geq 0$. We have $\Phi(0)=0$,
and
\begin{equation}
  \label{eq:Phi}
  \frac{d\Phi}{d\kappa} = 1 - \frac{ W(c_0 e^{\kappa})} { 1+ W(c_0
    e^{\kappa})} [1-e^{-\kappa}] - e^{-\kappa}W(c_0 e^{\kappa})\ ,
\end{equation}
i.e. $\Phi'(0)=1-W(c_0)$. The monotonous function $W(c_0)$ becomes
larger than 1 for $c_0>e$, i.e. in this case $p_1(t)$ would approach
zero from negative values. This is a contradiction. We therefore
conclude
\begin{equation}
  \label{eq:LRconlude}
  \forall c_0 < e: \qquad x_{LR}(c_0) = 1 - \frac{W(c_0)^2
    +2W(c_0)}{2c_0} \ ,
\end{equation}
and this value coincides with the relative size of a minimal vertex
cover. For $c_0>e$ the algorithm gets stuck if $p_1(t)=0$ is reached,
no vertices of degree one are left, and a finite fraction of all edges
remains uncovered.

\subsection{Generalized leaf removal}
\label{sec:leafgeneral}

In order to overcome this problem, we generalize the leaf-removal
algorithm by modifying the selection weights
to $w_d= 1+A \delta_{d,1}$. This algorithm interpolates between the
algorithms of Bauer and Golinelli ($A\to\infty$, exact minimal VCs for
$c_0<e$) and the one of Gazmuri ($A=0$, correct leading asymptotic 
behaviour for $c_0\gg 1$). We thus expect that this algorithm shows a
very good performance in the whole finite-connectivity region for
large, but finite $A$. 

\begin{figure}[htb]
\begin{center}
\includegraphics[width=15cm]{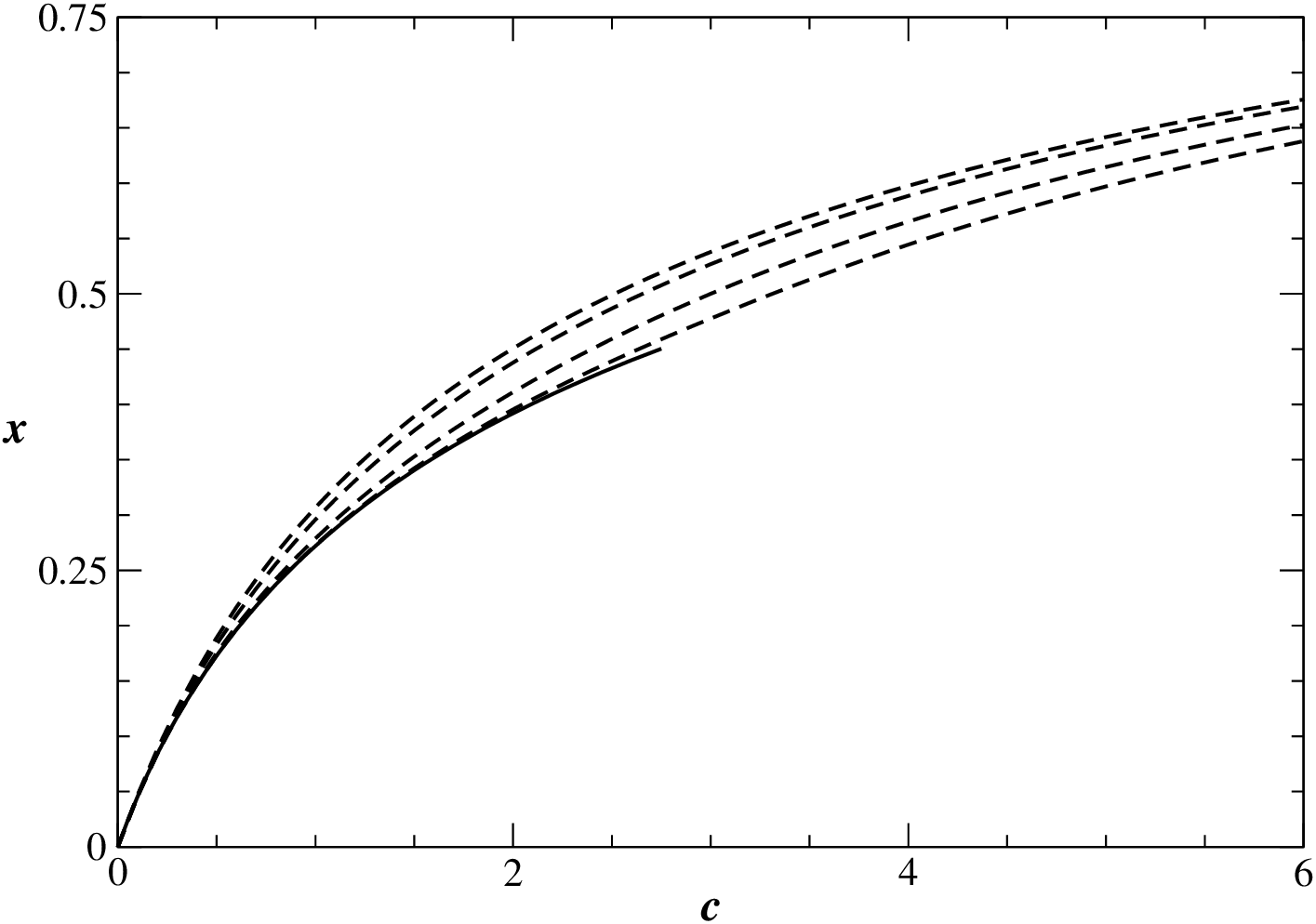}
\end{center}
\caption{Final size $x_f(c)$ of the vertex covers constructed by
  generalized leaf removal. The selection weight is
  $w_d=A\delta_{d,1}+1$ with $A=0,1,10,100$ (dashed lines from top to
  bottom). For a comparison, the result of the original leaf-removal
  algorithm is represented by the full line: For $c<e$, a minimal VC
  is found, whereas the algorithm fails completely to construct a VC
  for $c>e$.}
\label{fig:glr}
\end{figure}

Also in this case, the Poissonian shape of $p_d(t)$ remains correct for
all degrees $d>1$, and ansatz (\ref{eq:LRansatz}) together with the
normalization constraint (\ref{eq:norm}) remains valid. Plugging
everything into the dynamical equations (\ref{eq:dgl}), we directly
arrive at
\begin{eqnarray}
  \label{eq:GLRdgl}
  \dot n(t) &=& - \frac{1+c(t)+2A p_1(t)}{1+A p_1(t)} \nonumber\\
  n(t) \dot c(t) &=& c(t) \frac{1+c(t)+2A p_1(t)}{1+A p_1(t)}
     -2 \frac{[\gamma(t)\kappa(t)^2+c(t)][c(t)+Ap_1(t)]}{[1+Ap_1(t)]
       c(t)} \nonumber\\
  n(t) \dot \kappa(t) &=& -\kappa(t )\frac{[\gamma(t)\kappa(t)^2+c(t)]
       [c(t)+Ap_1(t)]}{[1+Ap_1(t)] c(t)^2} \nonumber\\
  n(t) \dot \gamma(t) &=& \gamma(t) \frac{c(t)+2A p_1(t)}{1+A p_1(t)} 
      -\kappa(t)\gamma(t)  \frac{c(t)+A p_1(t)}{1+A p_1(t)}
\end{eqnarray}
with $p_1(t) = c(t)-\gamma(t)\kappa(t)[1-e^{-\kappa(t)}]$. These
equations determine the exact graph reduction dynamics for generalized
leaf removal, and can be solved numerically. Due to the non-zero
selection weights for all degrees, these equations do not suffer from
the appearance of negative values for certain $p_d(t)$, and the
algorithm always constructs a vertex cover. The number of covered
vertices at algorithmic time $t$ can be calculated from
eq. (\ref{eq:xodd}), which for our special choices of $k$ and $w_d$
reads 
\begin{equation}
  \label{eq:GLRxdot}
  \dot x(t) = \frac{ c(t)+ Ap_1(t)}{1+Ap_1(t)}
\end{equation}
The relative size of the finally constructed vertex cover is, with
probability one, given by $x(t_f)$, with $t_f$ following from 
$c(t_f)=0$. In figure \ref{fig:glr}, the results are presented for
several values of $A$ and compared with Gazmuri's resp. Bauer and
Golinelli's algorithms.


\section{Summary and outlook}\label{sec:conclusion}

The solution of many combinatorial optimization problems requires
exponential time resources and is thus restricted to relatively small
system sizes. For larger systems, good and fast approximation
algorithms are needed, which are frequently based on heuristic
considerations concerning correlations between local problem structure
and optimal problem solutions.

Constructing a minimal vertex cover of a given graph belongs to the
basic NP-hard problems, and can be understood as a prototype
combinatorial optimization problem. In this paper, we have therefore
analysed linear-time algorithms for constructing small vertex covers
of finite-connectivity random graphs. The applied heuristic exploits
the observation that vertices of high vertex degree are more likely to
be covered in minimal VCs, whereas those of small degree remain more
frequently uncovered.

We have introduced and analysed mainly two types of algorithms, namely
depth-zero and depth-one algorithms. In the first case, vertices are
selected randomly and (if connected to any other vertex) they are
covered. We found that the performance of the algorithm can be largely 
improved by preferentially selecting vertices of high degree.

We also observed that depth-zero algorithms were outperformed by
depth-one algorithms. These select a vertex in every algorithmic step 
and uncover it. All neighbours of the selected vertex must be covered
consequently. The best performance is achieved if the algorithm
always selects a vertex of smallest degree. As was already found by
Bauer and Golinelli \cite{BaGo2}, this procedure even outputs an (almost)
minimal VC if it is applied to random graphs of average connectivity
$c_0<e$. The algorithm constructs good approximations also for higher
connectivities.

Both types of algorithms can be interpreted as Markovian graph
reduction processes. They are analytically characterized by the
evolution of the degree distribution of the remaining uncovered
subgraphs. The dynamical equations were solved in some cases, in other
cases approximations were necessary.

The presented approach can be extended into several directions:
\begin{itemize}
\item The applied heuristic was restricted to considering the simplest
  local structure, namely the degree of the selected vertex. Depth-one
  algorithms can be improved by e.g. selecting a vertex of minimal
  degree but with maximal number of next-nearest neighbours. Covering
  the nearest neighbours thus results in a higher number of covered
  edges. It would be interesting to extend the rate equations to this
  case.
\item VC is used as a prototype optimization problem, but the approach
  can be generalized to other combinatorial problems defined over
  random structures, e.g. to graph coloring or satisfiability
  problems. For random 3-satisfiability, lower bounds for the SAT/UNSAT
  threshold are usually obtained using algorithms in the so-called 
  card-game representation \cite{Fra,Ac}, which corresponds to
  $w_d=const$ in our analysis.
\item Our analysis was restricted to the typical time evolution of the
  degree distribution. Deviations appear with exponentially small
  probability - and are thus important for small systems. These rare
  events can be systematically exploited by exponentially frequent
  restarts of the algorithm. If a minimal VC is found with probability 
  $p=e^{-\tau N}$, we need $e^{(\tau+\varepsilon)N}$ restarts to
  almost surely construct a VC (for all $\varepsilon >0$). As observed 
  recently for simple heuristics of vertex cover \cite{MonZe} and
  3-satisfiability \cite{CoMo}, this random restart algorithm can be
  exponentially faster than exact standard procedures (like
  backtracking). The performance can be improved further by using a
  more sophisticated heuristics. 
\item Also sophisticated complete algorithms, i.e. those that find an
  optimal solution for sure, use heuristic arguments for accelerating
  the combinatorial search. The presented ideas may hence contribute
  to the analysis of such algorithms, and the insight may be used to
  exponentially speed up the numerical search.
\end{itemize}

\newpage

\newpage

\newpage

\begin{thebibliography}{99}
\bibitem{GaJo} M.R. Garey and D.S. Johnson, {\it Computers and
    Intractability}, (Freeman, San Francisco 1979).
\bibitem{note:ptas} Please note that ``approximation'' is not used in
  the sense of a polynomial-time approximation scheme, which
  guarantees to approximate the optimal solution within some given
  error for arbitrary samples. Here we mean low-cost configurations 
  which may serve as reasonable solutions in practical applications.
\bibitem{AI} special issue of Art. Int. {\bf 81} (1996).
\bibitem{TCS} special issue of Theor. Comp. Sci. {\bf 265} (2001).
\bibitem{Ga} P.G. Gazmuri, Networks {\bf 14}, 367 (1984).
\bibitem{Wo} N.C. Wormald, Ann. Appl. Prob. {\bf 5}, 1217 (1995).
\bibitem{PiSpWo} B.G. Pittel, J. Spencer, and N.J. Wormald,
  J. Combinat. Theor. B {\bf 67}, 111 (1996).
\bibitem{Fra} J. Franco, Theor. Comp. Sci. {\bf 265}, 147 (2001).
\bibitem{Ac} D. Achlioptas, Theor. Comp. Sci. {\bf 265}, 159 (2001).
\bibitem{MoZe} R. Monasson and R. Zecchina, Phys. Rev. E {\bf 56},
  1357 (1997).
\bibitem{nature} R. Monasson, R. Zecchina, S. Kirkpatrick, B. Selman,
  and L. Troyansky, Nature {\bf 400}, 133 (1999). 
\bibitem{Me} S. Mertens, Phys. Rev. Lett. {\bf 81}, 4281 (1998).
\bibitem{WeHa1} M. Weigt and A.K. Hartmann, Phys. Rev. Lett. {\bf 84},
  6118 (2000)
\bibitem{WeHa2} A.K. Hartmann and M. Weigt, Theor. Comp. Sci. {\bf
    265}, 199 (2001).
\bibitem{BaGo1} M. Bauer and O. Golinelli, Phys. Rev. Lett. {\bf 86},
  2621 (2001).
\bibitem{BaGo2} M. Bauer and O. Golinelli, Eur. Phys. J. B {\bf 24},
  339 (2001).
\bibitem{WeHa3} M. Weigt and A.K. Hartmann, Phys. Rev. E {\bf 63},
  056127 (2001).
\bibitem{WeHa4} M. Weigt and A.K. Hartmann, Phys. Rev. Lett. {\bf 86},
  1658 (2001).
\bibitem{note1} This works at least for the case where the graph
  contains no loops of length up to $2k$, which is allways true for 
  the most inportant cases $k=0,1$. Also for random graphs,
  $G^{(k)}(i)$ is almost allways a tree, and the above argument works
  with probability one (for $N\to\infty$).
\bibitem{Bo} B. Bollob\'as, {\it Random Graphs}, (Academic Press 1985).
\bibitem{ErRe} P. Erd\"os and A. R\'enyi,
  Publ. Math. Inst. Hung. Acad. Sci. {\bf 5}, 17 (1960).
\bibitem{MoRe} M. Molloy and B. Reed, Rand. Struct. Alg. {\bf 6}, 161
  (1995). 
\bibitem{Ne} M.E.J. Newman, S.H. Strogatz, and D.J. Watts,
  Phys. Rev. E {\bf 64}, 026118 (2001).
\bibitem{AlBa} R. Albert and A. L. Barabasi, Rev. Mod. Phys. 74, 47
  (2002).  
\bibitem{KrRe} P.L. Krapivsky, S. Redner, and F. Leyvraz,
  Phys. Rev. Lett. {\bf 85}, 4629 (2000).
\bibitem{DoMe} S.N. Dorogovtsev, J.F.F. Mendes, and A.N. Samukhin,
  Phys. Rev. Lett. {\bf 85}, 4633 (2000).
\bibitem{Fr} A.M. Frieze, Discrete Math. {\bf 81}, 171 (1990).
\bibitem{MonZe} A. Montanari and R. Zecchina, Phys. Rev. Lett. {\bf
    88}, 178701 (2002).
\bibitem{CoMo} S. Cocco and R. Monasson, preprint cond-mat/0203012.
\end{thebibliography}
\end{document}